\documentstyle[12pt]{article}    
\newcommand{\eq}{\begin{eqnarray}}
\newcommand{\en}{\end{eqnarray}}
\begin{document}
\thispagestyle{empty}

\hfill {\bf BUTP-99/25}

\vspace*{.3cm}
\begin{center}

{\Large\bf Spectrum and Decays of Hadronic Atoms}

\vspace*{.5cm}
J. Gasser\\
{\em Institute for Theoretical Physics, University of Bern,
Sidlerstrasse 5, CH-3012, Bern, Switzerland}\\
\vspace{0.3cm}
V. E. Lyubovitskij\footnote{Present address: 
Institute of Theoretical Physics, University of T\"{u}bingen, 
Auf der Morgenstelle 14, D-72076 T\"{u}bingen, Germany} \\
{\em Bogoliubov Laboratory of Theoretical Physics, Joint Institute
for Nuclear Research, 141980 Dubna, Russia and\\
Department of Physics, Tomsk State University, 
634050 Tomsk, Russia}\\
\vspace{0.3cm}
and \\
\vspace{.3cm}
\underline{A. Rusetsky}\\
{\em Institute for Theoretical Physics, University of Bern,
Sidlerstrasse 5, CH-3012, Bern, Switzerland, \\
Bogoliubov Laboratory of Theoretical Physics, Joint Institute
for Nuclear Research, 141980 Dubna, Russia and\\
HEPI, Tbilisi State University, 380086 Tbilisi, Georgia}

\end{center}

\begin{abstract}
\setlength{\baselineskip}{2.6ex}

\noindent
 Using non relativistic effective Lagrangian techniques,
 we analyze the hadronic decay of the $\pi^+\pi^-$ atom and the
 strong energy-level shift of  pionic hydrogen in the ground state.
 We derive  general formulae for the width and level shift,
 valid  at next-to-leading order in isospin breaking.
 The result is expressed  in terms of  hadronic threshold amplitudes
 that   include isospin-breaking effects.
 In order to extract isospin symmetric  scattering lengths from  
 the data, we invoke chiral perturbation theory, that allows one to relate
 the scattering lengths to the threshold amplitudes. 

\end{abstract}

\newpage

Recent years have seen a  growing  interest in the 
study of  hadronic atoms. At CERN,
 the DIRAC collaboration~\cite{DIRAC}
 aims to measure the $\pi^+\pi^-$ atom lifetime 
 to $10\%$ accuracy. This would allow one to determine the difference 
 $a_0-a_2$ of $\pi\pi$ scattering lengths with  $5\%$ precision.
This measurement provides a crucial test for the large/small
condensate scenario in QCD: should it turn out that the quantity $a_0-a_2$ is 
different from the value predicted in standard ChPT~\cite{ChPT}, one has to 
conclude~\cite{Stern} 
that spontaneous  chiral symmetry breaking in QCD proceeds differently from 
the widely accepted picture~\cite{GOR}. In the experiment performed at 
PSI~\cite{PSI1}, one has measured the strong energy-level shift
 and the total decay width of the $1s$ state of pionic hydrogen, as well as
 the $1s$ shift of pionic deuterium.
Using the technique described in 
Ref.~\cite{PSI1}, these measurements  yield ~\cite{PSI2} 
isospin symmetric  $\pi N$ scattering
lengths to an 
accuracy which is unique for hadron physics: 
$a_{0+}^+=(1.6\pm 1.3)\times 10^{-3}M_{\pi^+}^{-1}$ and
$a_{0+}^- = (86.8\pm 1.4)\times 10^{-3} M_{\pi^+}^{-1}$.
 The scattering length $a_{0+}^-$  may be  used as an input
in the Goldberger-Miyazawa-Oehme~\cite{GMO} 
sum rule to determine the $\pi NN$ coupling 
constant~\cite{PSI1,PSI2}.
A new experiment 
on  pionic hydrogen~\cite{Gotta} has recently been  approved. It will
allow one to measure the decay $A_{\pi^-p}\rightarrow \pi^0n$  to much
higher accuracy and thus enable one, 
in principle, to determine
the $\pi N$ scattering lengths from  data on pionic hydrogen alone.
This might vastly reduce the model-dependent 
uncertainties that come from the analysis of the three-body problem in
$A_{\pi^-d}$.
 Finally, the DEAR collaboration~\cite{DEAR} 
 at the DA$\Phi$NE facility  (Frascati) plans to measure  
the energy level shift and lifetime of the $1s$ state in $K^{-}p$ 
 and $K^-d$ atoms
 - with considerably higher precision 
than in the previous experiment carried out at KEK~\cite{KEK} for
$K^-p$ atoms. It is expected~\cite{DEAR}
 that this will result in a precise determination of the $I=0,1$ $S$-wave
 scattering lengths - although, of course, one will  again be faced 
with the three-body problem already mentioned. It will be a challenge for 
theorists to
extract from this new information on the $\bar{K}N$ amplitude at threshold 
a more precise value of e.g. the isoscalar 
kaon-sigma term and of
the strangeness content of the nucleon~\cite{Gensini}.

 We now turn to    {\it theoretical} investigations of hadronic atoms.
At leading order in isospin breaking, the  
energy-level shift and the  decay width of these atoms
  can be expressed in terms of the 
strong hadronic scattering lengths through the well-known formulae by Deser
{\it et al.}~\cite{Deser}. More precisely, these formulae relate the
  ground state level
shift - induced by the strong interaction -
and  its partial
 decay width into neutral hadrons (e.g., 
$A_{\pi^+\pi^-}\rightarrow\pi^0\pi^0$, $A_{\pi^- p}\rightarrow\pi^0 n$)
to the corresponding isospin combinations of strong scattering lengths,
\begin{eqnarray}\label{deser}
\Delta E_{{\rm str}}\sim\Psi_0^2\, {\rm Re}\, a_{cc}\, ,\quad\quad
\Gamma_{c0}\sim({\rm phase~space})\times \Psi_0^2\, |a_{c0}|^2\, .
\end{eqnarray}
Here, $\Psi_0$ denotes the value of the Coulomb wave function at the origin,
 and
  $a_{cc}$, $a_{c0}$ stand for 
the relevant isospin combinations of strong
scattering lengths. We have used the notation "$c$" for "charged" 
(e.g., $\pi^+\pi^-$, $\pi^- p$) and "$0$" for "neutral" (e.g., $\pi^0\pi^0$,
$\pi^0 n$) channels. 
The accuracy of these leading-order
formulae  is however not sufficient to fully exploit 
existing and forthcoming high-precision
 data  on hadronic atoms. Indeed, for that purpose, 
 one has to evaluate isospin-breaking 
 corrections  at next-to-leading  order.
 The aim of the present talk is to show how this can be achieved.

Recently, using a  non relativistic effective Lagrangian framework,
a general expression for the decay width $\Gamma_{A_{2\pi}\to\pi^0\pi^0}$
 of the $1s$ state of the 
$\pi^+\pi^-$ atom  was obtained at  next-to-leading order in
isospin-breaking~\cite{Bern}. We denote the 
fine-structure constant $\alpha$ and the quark mass
difference squared $(m_d-m_u)^2$  by the common symbol $\delta$.
Then, the decay width is written in the following form\footnote{
We use throughout the Landau symbols $O(x)$ [$o(x)$] for quantities that
vanish like $x$ [faster than $x$] when $x$ tends to zero. Furthermore,
it is understood that this holds modulo logarithmic terms, i.e. we write also
$O(x)$ for $x\ln x$.},
\begin{eqnarray}\label{general-pipi}
\Gamma_{A_{2\pi}\to\pi^0\pi^0}&=&\frac{2}{9}\,\alpha^3 p^\star 
{\cal A}_{\pi\pi}^{~2} (1+K_{\pi\pi})\, , \quad\quad 
{\cal A}_{\pi\pi}=a_0-a_2+O(\delta)\, , 
\nonumber\\[2mm]
K_{\pi\pi}&=&
\frac{\Delta M_\pi^2}{9M^2_{\pi^+}}\,(a_0+2a_2)^2
-\frac{2\alpha}{3}\,(\ln\alpha-1)\,(2a_0+a_2)+o(\delta)\, . 
\end{eqnarray}
Here $p^\star=(M_{\pi^+}^2-M_{\pi^0}^2-\frac{1}{4}M_{\pi^+}^2\alpha^2)^{1/2}$,
 and $a_I~(I=0,2)$ denote the strong $\pi\pi$ scattering lengths
in the channel with total isospin $I$, and the quantity ${\cal A}_{\pi\pi}$ is
calculated as follows~\cite{Bern}. One calculates the relativistic amplitude
for the process $\pi^+\pi^-\rightarrow\pi^0\pi^0$ at $O(\delta)$ in the
normalization chosen so that at $O(\delta^0)$ the amplitude at threshold 
coincides with the difference $a_0-a_2$ of (dimensionless) 
$S$-wave $\pi\pi$ scattering lengths. Due to the presence of virtual photons,
the amplitude is multiplied by an overall Coulomb phase that is removed.
The real part of the remainder contains  terms
that diverge like $|{\bf p}|^{-1}$ and $\ln 2|{\bf p}|/M_{\pi^+}$ at
$|{\bf p}|\rightarrow 0$ (${\bf p}$ denotes the relative 3-momentum
of charged pion pairs). 
 The quantity ${\cal A}_{\pi\pi}$
 is obtained by subtracting these divergent pieces, and by then evaluating the
remainder at ${\bf p}=0$. We shall refer to 
 ${\cal A}_{\pi\pi}$ as the physical scattering amplitude at threshold.

A few remarks are in order. As it is seen explicitly from 
Eq.~(\ref{general-pipi}), one can directly extract the value of 
${\cal A}_{\pi\pi}$
from the measurement of the  decay width, because
the correction  $K_{\pi\pi}$ 
is very small and the error introduced by it is
negligible.
 We emphasize that in derivation of Eq.~(\ref{general-pipi}), chiral
expansions have not been used.
 On the other hand, if one further aims to extract strong scattering
lengths from  data, one may invoke chiral perturbation theory (ChPT)
and to relate the quantities ${\cal A}_{\pi\pi}$ and
$a_0-a_2$ order by order in the chiral expansion. This requires
the evaluation of isospin-breaking corrections to the scattering amplitude.

The corrections to the hadronic atom characteristics, evaluated in this manner
contain, in general, contributions which have not been taken into 
account so far within the potential scattering  approach to the same 
problem~\cite{PSI1,Rasche}. An obvious candidate for these contributions is the
effect coming from the direct quark-photon coupling that is encoded in the
so-called "electromagnetic" low-energy constants (LEC's) in ChPT.
A second effect  is related 
 to the convention-dependent definition of the isospin-symmetric world
against which the isospin-breaking corrections are calculated. We adopt the
widely used convention that the masses of the isospin multiplets
$(\pi^\pm,\pi^0)$ and $(p,n)$ in this world
coincide with the masses of the charged particles in the real world.
This definition induces a contribution to the isospin-breaking corrections in
 the level shifts and decay widths.
 We shall display below both corrections explicitly
in the case of the $\pi^- p$  energy-level shift, where these effects emerge
already at tree level.

The investigation of the $\pi^-p$ atom  is 
very similar to the procedure used in the description of  the $\pi^+\pi^-$
 atom~\cite{Bern}.
In the following, we restrict
ourselves to the case of the strong energy-level shift of the $\pi^- p$ 
atom in the ground state.
Because the proton-neutron mass difference contains terms linear in $m_d-m_u$,
we count $\alpha$ and
$m_d-m_u$ as quantities of the same order, and denote them by the common
symbol $\delta'$. [Since this counting is merely a matter of convenience,
our previous results on the $\pi^+\pi^-$ atom remain of course unaltered.]
Further, for the energy shift of hadronic atoms, one
can no longer neglect the electromagnetic contributions coming from transverse
photons as it was done in the case of the width of the $\pi^+\pi^-$ atom. 
The reason for this
can easily be seen from counting  powers of $\alpha$ in the energy-level
shift. The binding energy of the atom starts at $O(\alpha^2)$
(nonrelativistic value $E_{{\rm NR}}=-\frac{1}{2}\, \mu_c\alpha^2$, 
where 
$\mu_c$ denotes the reduced mass of $\pi^- p$ system), 
and the corresponding QED corrections  start at $O(\alpha^4)$.
According to  Eq.~(\ref{deser}), the leading-order strong energy-level
shift is  O($\alpha^3$), while the next-to-leading  order corrections 
start at $O(\alpha^4)$ and should therefore be treated on the same footing 
as the QED
corrections\footnote{
There is one important exception to this rule. Though the vacuum polarization
correction starts at $O(\alpha^5)$, it is amplified by a large factor 
$(\mu_c/m_e)^2$, where $m_e$ denotes the electron mass. Since
$\alpha\mu_c/m_e\sim 1$, this contribution 
is numerically as important as the leading-order strong 
contribution (see~\cite{PSI1}). The graph responsible for this contribution can
be, however, easily singled out and the contribution from it merely added 
to the final result.}.
QED corrections, however, are not considered here - we focus on the strong
energy-level shift alone. For the latter, it is straightforward to obtain a
general formula very similar to Eq.~(\ref{general-pipi}), that gives the
strong energy-level shift including $O(\delta')$ corrections:
\begin{eqnarray}\label{general-piN}
\Delta E_{{\rm str}}=-2\alpha^3\mu_c^2\, {\cal A}_{\pi N}\,(1+K_{\pi N})\,,
\end{eqnarray}
where $K_{\pi N}$ is a quantity of order $\delta'$ (modulo logarithms) and can
be expressed in terms of the $S$-wave $\pi N$ scattering lengths $a_{0+}^+$ and
$a_{0+}^-$. Since $K_{\pi N}$ is small, the error introduced by the uncertainty in 
the determination of $a_{0+}^+,~a_{0+}^-$ is negligible. The major uncertainty in
the energy-level shift  comes from the quantity ${\cal A}_{\pi N}$ whose 
definition
is very similar to that of ${\cal A}_{\pi\pi}$. To evaluate this quantity,
one has to calculate the relativistic scattering amplitude for the process
$\pi^-p\rightarrow\pi^-p$ at $O(\delta')$, subtract all diagrams that are
made disconnected by cutting one virtual photon line and remove the Coulomb
phase. The real part of the remainder, as for the $\pi^+\pi^-$ case,   
contains singular pieces that behave like $|{\bf p}|^{-1}$ and
$\ln|{\bf p}|/\mu_c$ that should be again subtracted 
(${\bf p}$ denotes the relative
3-momentum of the $\pi^- p$ pair in CM). The rest - evaluated at ${\bf p}=0$ -
coincides, by definition, with ${\cal A}_{\pi N}$. [The normalization of the
relativistic amplitude is chosen so that 
${\cal A}_{\pi N}=a_{0+}^+ + a_{0+}^-+O(\delta')$.]
  
Further, to analyze the isospin-breaking corrections to the 
energy-level shift, we relate the 
physical scattering amplitude at threshold ${\cal A}_{\pi N}$ to the
scattering lengths $a_{0+}^+,~a_{0+}^-$ in ChPT. 
At $O(p^2)$ in the chiral expansion, where the amplitude
is determined by tree diagrams, this relation is remarkably simple.
 Constructed on the basis of the effective $\pi N$ 
Lagrangian~\cite{GSS,MS,becher},
 the amplitude contains the pseudovector Born term 
 ${\cal A}_{\pi N}^{{\rm pv}}$ with physical masses, and a 
 contribution that contains a linear combination of 
 $O(p^2)$ LEC's, 
\begin{eqnarray}\label{amplitude}
{\cal A}_{\pi N}^{(2)}&=&a_{0+}^+ + a_{0+}^-
 +\epsilon_{\pi N}^{(2)}\nonumber\\
&=&{\cal A}_{\pi N}^{{\rm pv}}+2\hat m B \kappa_1c_1
+M_{\pi^+}^2 (\kappa_2c_2+\kappa_3c_3)+e^2 (\sigma_1f_1+\sigma_2f_2)\, ,
\end{eqnarray}
where the quantity $B$ is related to the quark
condensate, and where $c_i$ ($f_i$) are strong  (electromagnetic)
LEC's
 from the $O(p^2)$ Lagrangian of ChPT. Furthermore, $\kappa_i$ and $\sigma_i$
denote isospin symmetric coefficients whose explicit expressions are not
needed here. From Eq.~(\ref{amplitude}), 
it is straightforward to visualize both mechanisms
of isospin-breaking corrections to the hadronic atom observables, not
included in potential approaches. The direct quark-photon 
coupling is encoded in
 the coupling constants $f_i$, whereas the effect of the  mass tuning in the
hadronic amplitude (described above) is due to
the term proportional to $2\hat{m}B$. Indeed, at this order in the chiral
expansion, one has $2\hat{m} B= M_{\pi^0}^2$. As we 
express the strong amplitude
in terms of charged masses by convention, we write
\begin{eqnarray}
2\hat{m}B=M_{\pi^+}^2 -\Delta_\pi\, ; \quad \Delta_\pi=  
M_{\pi^+}^2-M_{\pi^0}^2\, ,
\end{eqnarray}
and obtain 
\begin{eqnarray}\label{epsilon}
\epsilon_{\pi N}^{(2)}=
-\Delta_\pi\, \kappa_1c_1+e^2\, (\sigma_1f_1+\sigma_2f_2)
+O(\hat m\delta')+o(\delta')\, .
\end{eqnarray}
Estimates for the 
energy-level shift on the basis of the expression~(\ref{epsilon})
will be presented elsewhere.
 Here we note that a simple  order-of-magnitude estimate for $f_1$
 shows that  $f_1$ induces an uncertainty 
in the energy-level shift of roughly the same size as the total correction
given in Ref.~\cite{PSI1}.

To summarize, we have applied a non relativistic effective Lagrangian approach
to the study of $\pi^+\pi^-$ and $\pi^- p$ atoms in the ground state.
A general expression for the width $\Gamma_{A_{2\pi}\to\pi^0\pi^0}$
and for the strong level shift of pionic hydrogen has been obtained at 
next-to-leading order in isospin breaking. The sources of the 
isospin-breaking corrections in these quantities, complementary to ones 
already considered in the potential scattering theory approach, have been 
clearly identified.

{\it Acknowledgments}. 
V. E. L. thanks the Organizing Committee of MENU99 Symposium for financial 
support and the University of Bern, where this work was performed, for 
hospitality.
This work was supported in part by the Swiss National Science
Foundation, and by TMR, BBW-Contract No. 97.0131  and  EC-Contract
No. ERBFMRX-CT980169 (EURODA$\Phi$NE).

\bibliographystyle{unsrt}

\end{document}